\documentstyle[12pt]{article}
\begin{document}
\begin{titlepage}
\begin{flushright}
IC/2001/90\\
hep-th/0108073
\end{flushright}
\vspace{10 mm}

\begin{center}
{\Large A Scalar-Tensor Bimetric Brane World Cosmology}

\vspace{5mm}

\end{center}
\vspace{5 mm}

\begin{center}
{\large Donam Youm\footnote{E-mail: youmd@ictp.trieste.it}}

\vspace{3mm}

ICTP, Strada Costiera 11, 34014 Trieste, Italy

\end{center}

\vspace{1cm}

\begin{center}
{\large Abstract}
\end{center}

\noindent

We study a scalar-tensor bimetric cosmology in the Randall-Sundrum model 
with one positive tension brane, where the biscalar field is assumed to be 
confined on the brane.  The effective Friedmann equations on the brane are 
obtained and analyzed.  We comment on resolution of cosmological problems 
in this bimetric model.

\vspace{1cm}
\begin{flushleft}
August, 2001
\end{flushleft}
\end{titlepage}
\newpage

Variable-Speed-of-Light (VSL) cosmologies were proposed \cite{mof1,am} as 
possible solutions to the initial value problems in the Standard Big Bang 
(SBB) model.  VSL models assume that the speed of light initially took a 
larger value and then decreased to the present day value at an early time.  
The VSL models solve various cosmological problems of the SBB model 
associated with the initial value problems, including those solved by the 
inflationary scenario \cite{gut,lin,als}.  In the original models by Moffat 
\cite{mof1} and by Albrecht and Magueijo \cite{am}, the speed of light $c$ 
(and possibly the Newton's constant $G_4$) in the action, which is a 
fundamental constant of the nature, is just assumed to vary with time during 
an early period of cosmic evolution and thereby the Lorentz symmetry becomes 
explicitly broken.  So, it becomes necessary to assume that there exists a 
preferred frame in which the laws of physics take simple forms.  Later, 
Clayton and Moffat \cite{cm1,cm2,cm3,cm4} proposed an ingenious mechanism 
by which the speed of light can vary with time in a diffeomorphism invariant 
manner and without explicitly breaking the Lorentz symmetry.  (See also Ref. 
\cite{dru} for an independent development.)  In their models, two metrics 
are introduced into the spacetime manifold (thereby their models are called 
bimetric), one being associated with gravitons and the other providing the 
geometry on which matter fields, including photons, propagate.  Since these 
two metrics are nonconformally related by a scalar field (called a biscalar) 
or a vector field (called a bivector), photons and gravitons propagate at 
different speeds.  

It has been shown \cite{kal1,kir,kal2,chu,ale,ish,ckr,csa} that brane world 
models manifest the Lorentz violation, which is a necessary requirement for 
the VSL models.  Therefore, it would be of interesting to study the VSL 
cosmologies within the context of the brane world scenarios.  In particular,  
the VSL models may provide a possible mechanism for bringing the quantum 
corrections to the fine-tuned brane tensions under control, since the VSL 
models generally solve the cosmological constant problem.  In our previous 
work \cite{youm}, we studied the VSL cosmologies in the Randall-Sundrum (RS) 
scenarios \cite{rs1,rs2}, following the approach of the earlier VSL models 
\cite{mof1,am,bar1,bar2,bar3,mof2,bar4} with varying fundamental constants.  
In this paper, we follow the approach of Clayton and Moffat to study the 
bimetric cosmology in the RS2 model \cite{rs2}.  

In the bimetric models it is usually assumed that gravitons and the biscalar
\footnote{As shown in Ref. \cite{cm2}, one can also rewrite the biscalar 
equation of motion in such a way that the biscalar field appears to 
propagate on a different geometry described by the metric expressed in terms 
of the gravity metric and the matter field energy-momentum tensor.} 
(or the bivector) propagate on the geometry described by the ``gravity 
metric'', whereas all the matter fields propagate on the geometry 
described by the ``matter metric''.  So, a natural bimetric modification of 
a brane world model would be just to modify the brane matter field action to 
be constructed out of the matter metric.  We consider the bimetric model with 
a biscalar, which is assumed to be confined on the brane worldvolume.  The 
action for the bimetric RS2 model with the brane matter fields therefore 
takes the following form:
\begin{eqnarray}
S=\int d^5x\sqrt{-G}\left[{c^4\over{16\pi G_5}}{\cal R}-\Lambda\right]
+\int d^4x\sqrt{-\hat{g}}{\cal L}_{\rm mat}
\cr
-\int d^4x\sqrt{-g}\left[{1\over 2}g^{\mu\nu}\partial_{\mu}\Phi
\partial_{\nu}\Phi+V(\Phi)+\sigma\right],
\label{bimetact}
\end{eqnarray}
where $\Phi$ is the biscalar field with the potential $V(\Phi)$ and $\sigma$ 
is the tension of the brane assumed to be located at the origin $y=0$ of the 
extra spatial coordinate $y$.  Here, the gravity metric $g_{\mu\nu}$ and the 
matter metric $\hat{g}_{\mu\nu}$ on the brane are given in terms of the bulk 
metric $G_{MN}$ and $\Phi$ by
\begin{equation}
g_{\mu\nu}=G_{\mu\nu}(x^{\rho},0),\ \ \ \ \ \ \ \ \ 
\hat{g}_{\mu\nu}=g_{\mu\nu}-B\partial_{\mu}\Phi\partial_{\nu}\Phi,
\label{matmet}
\end{equation}
where a dimensionless constant $B$ is assumed to be positive.  The Lagrangian 
${\cal L}_{\rm mat}$ for the brane matter fields is constructed out of 
$\hat{g}_{\mu\nu}$.  

We study the brane world cosmology associated with the above action.  
The general metric ansatz for the expanding brane universe where the 
principle of homogeneity and isotropy in the three-dimensional space 
on the three-brane is satisfied is given by
\begin{equation}
G_{MN}dx^Mdx^N=-n^2(t,y)c^2dt^2+a^2(t,y)\gamma_{ij}dx^idx^j+b^2(t,y)dy^2,
\label{blkmet}
\end{equation}
where $\gamma_{ij}$ is the metric for the maximally symmetric 
three-dimensional space given in the Cartesian and the spherical 
coordinates by
\begin{equation}
\gamma_{ij}dx^idx^j=\left(1+\textstyle{k\over 4}\delta_{mn}x^mx^n
\right)^{-2}\delta_{ij}dx^idx^j={{dr^2}\over{1-kr^2}}+r^2(d\theta^2+
\sin^2\theta d\phi^2),
\label{mxsymet}
\end{equation}
with $k=-1,0,1$ respectively for the three-dimensional spaces with the 
negative, zero and positive spatial curvatures.  Making use of the fact 
that it is always possible to choose a gauge so that $n(t,0)$ is constant 
without introducing the cross term $G_{ty}$, we scale the time coordinate 
such that $n(t,0)=1$.  With the assumption of homogeneity and isotropy on 
the three-brane, the biscalar field $\Phi$ does not depend on the spatial 
coordinates $x^i$ ($i=1,2,3$) of the three-brane.  

In obtaining the Einstein's equations by varying the action w.r.t. the metric,
we have to keep in mind that $\hat{g}_{\mu\nu}$ is the physical metric for 
matter fields on the brane.  So, the energy-momentum tensor for the matter 
fields are defined in terms of $\hat{g}_{\mu\nu}$:
\begin{equation}
\hat{T}^{\mu\nu}\equiv{2\over\sqrt{-\hat{g}}}{{\delta(\sqrt{-\hat{g}}
{\cal L}_{\rm mat})}\over{\delta\hat{g}_{\mu\nu}}}.
\label{matemtens}
\end{equation}
Modeling the brane matter fields as perfect-fluid, we can put this 
energy-momentum tensor into the following standard form:
\begin{equation}
\hat{T}^{\mu\nu}=\left(\varrho+{\wp\over c^2}\right)U^{\mu}U^{\nu}+
\wp\hat{g}^{\mu\nu},
\label{pfemtens}
\end{equation}
where $\varrho$, $\wp$ and $U^{\mu}$ are respectively the mass 
density, the pressure and the four-velocity of the fluid, and the 
inverse $\hat{g}^{\mu\nu}$ of $\hat{g}_{\mu\nu}$ is given by  
\begin{equation}
\hat{g}^{\mu\nu}=g^{\mu\nu}+{B\over{1-Bg^{\alpha\beta}\partial_{\alpha}\Phi
\partial_{\beta}\Phi}}g^{\mu\mu{\prime}}g^{\nu\nu^{\prime}}
\partial_{\mu^{\prime}}\Phi\partial_{\nu^{\prime}}\Phi.
\label{invmatmet}
\end{equation}
The four-velocity is normalized as $\hat{g}_{\mu\nu}U^{\mu}U^{\nu}=-c^2$.  
So, in the comoving coordinates, the nonzero component of $U^{\mu}$ is 
given by
\begin{equation}
U^t={1\over\sqrt{1+B\dot{\Phi}^2/c^2}},
\label{nzrumu}
\end{equation}
where the overdot denotes derivative w.r.t. $t$.  
The nonzero components of the energy-momentum tensor for the brane matter 
fields are therefore
\begin{equation}
\hat{T}^{tt}={\varrho\over{1+B\dot{\Phi}^2/c^2}},\ \ \ \ \ \ \ \ \ \ \ 
\hat{T}^{ij}={\wp\over a^2_0}\gamma^{ij}.
\label{nnzremtns}
\end{equation}
Since we assume that the matter field action is constructed out of 
$\hat{g}_{\mu\nu}$, the equations of motion of the brane matter fields imply 
the conservation law for the brane matter fields energy-momentum tensor:
\begin{equation}
{1\over\sqrt{-\hat{g}}}\partial_{\mu}\left(\sqrt{-\hat{g}}\hat{T}^{\mu\nu}
\right)=0,
\label{conslaw}
\end{equation}
which takes the following standard form after the above expressions for the 
energy-momentum tensor and the matter metric are substituted:
\begin{equation}
\dot{\varrho}+3\left(\varrho+{\wp\over c^2}\right){\dot{a}_0\over a_0}=0,
\label{cnsvlaw}
\end{equation}
where the subscript 0 denotes quantities evaluated at $y=0$, i.e., 
$a_0(t)\equiv a(t,0)$.  This conservation equation can be derived 
independently from the effective four-dimensional Friedmann equations 
on the brane, implying that Eq. (\ref{conslaw}) is consistent with 
the Einstein's equations (\ref{eineq}) in the below.  

Taking the variation of the action $S$ w.r.t. the metric, we obtain the 
following Einstein's equations:
\begin{equation}
{\cal G}^{MN}={{8\pi G_5}\over c^4}{\cal T}^{MN},
\label{eineq}
\end{equation}
with the total energy-momentum tensor given by
\begin{eqnarray}
{\cal T}^{MN}&=&-G^{MN}\Lambda+\delta^M_{\mu}\delta^N_{\nu}\left[
\hat{T}^{\mu\nu}{\sqrt{-\hat{g}}\over\sqrt{-G}}+
\left\{g^{\mu\mu^{\prime}}g^{\nu\nu^{\prime}}\partial_{\mu^{\prime}}\Phi
\partial_{\nu^{\prime}}\Phi-\textstyle{1\over 2}g^{\mu\nu}\partial_{\alpha}
\Phi\partial^{\alpha}\Phi\right.\right.
\cr
& &\left.\left.-g^{\mu\nu}V(\Phi)-g^{\mu\nu}\sigma\right\}{\sqrt{-g}
\over\sqrt{-G}}\right]\delta(y).
\label{emtens}
\end{eqnarray}  
The equation of motion for the biscalar is
\begin{equation}
\nabla^2\Phi-V^{\prime}(\Phi)+B{\sqrt{-\hat{g}}\over\sqrt{-g}}\hat{T}^{\mu\nu}
\hat{\nabla}_{\mu}\hat{\nabla}_{\nu}\Phi=0,
\label{biscleq}
\end{equation}
where $\nabla^2\Phi=g^{\mu\nu}\nabla_{\mu}\nabla_{\nu}\Phi$ and 
the covariant derivative $\nabla_{\mu}$ [$\hat{\nabla}_{\mu}$] is 
defined in terms of the metric $g_{\mu\nu}$ [$\hat{g}_{\mu\nu}$].  We made 
use of the conservation law (\ref{conslaw}) to achieve the simplified form
of the last term on the LHS and the prime on the biscalar potential $V$ 
denotes derivative w.r.t. $\Phi$.  The above biscalar equation has dependence 
on $\hat{T}^{\mu\nu}$ due to the fact that the matter metric 
$\hat{g}_{\mu\nu}$, of which the brane matter fields action is made, depends 
on $\partial_{\mu}\Phi$.   

After the above ansatz for the metric and the biscalar field are substituted, 
the Einstein's equations (\ref{eineq}) take the following forms:
\begin{eqnarray}
{3\over{n^2c^2}}{\dot{a}\over a}\left({\dot{a}\over a}+{\dot{b}\over b}\right)
-{3\over b^2}\left[{a^{\prime}\over a}\left({a^{\prime}\over a}-{b^{\prime}
\over b}\right)+{a^{\prime\prime}\over a}\right]+{{3k}\over a^2}=
{{8\pi G_5}\over c^4}\left[\Lambda+\left(\varrho_{\Phi}c^2+{{\varrho 
c^2}\over\sqrt{I}}+\sigma\right){{\delta(y)}\over b}\right],
\label{ttein}
\end{eqnarray}
\begin{eqnarray}
{1\over b^2}\left[{a^{\prime}\over a}\left(2{n^{\prime}\over n}+{a^{\prime}
\over a}\right)-{b^{\prime}\over b}\left({n^{\prime}\over n}+2{a^{\prime}
\over a}\right)+2{a^{\prime\prime}\over a}+{n^{\prime\prime}\over n}\right]
\ \ \ \ \ \ \ \ \ \ \ \ \ \ \ \ \ \ \ 
\cr
+{1\over{n^2c^2}}\left[{\dot{a}\over a}\left(2{\dot{n}\over n}-{\dot{a}\over 
a}\right)+{\dot{b}\over b}\left({\dot{n}\over n}-2{\dot{a}\over a}\right)-
2{\ddot{a}\over a}-{\ddot{b}\over b}\right]-{k\over a^2}=
\ \ \ 
\cr
{{8\pi G_5}\over c^4}\left[-\Lambda+\left(\wp_{\Phi}+\sqrt{I}\wp-\sigma
\right){{\delta(y)}\over b}\right],
\ \ \ \ \ \ \ \ \ \ \ \ \ \ \ \ \ \ \ \ \ \ \ 
\label{iiein}
\end{eqnarray}
\begin{equation}
{n^{\prime}\over n}{\dot{a}\over a}+{a^{\prime}\over a}{\dot{b}\over b}
-{\dot{a}^{\prime}\over a}=0,
\label{tyein}
\end{equation}
\begin{equation}
{3\over b^2}{a^{\prime}\over a}\left({a^{\prime}\over a}+{n^{\prime}\over n}
\right)-{3\over{n^2c^2}}\left[{\dot{a}\over a}\left({\dot{a}\over a}-{\dot{n}
\over n}\right)+{\ddot{a}\over a}\right]-{{3k}\over a^2}=
-{{8\pi G_5}\over c^4}\Lambda,
\label{yyein}
\end{equation}
where the primes on the metric components denote derivative w.r.t. $y$.  
Here, the biscalar field mass density $\varrho_{\Phi}$ and pressure 
$\wp_{\Phi}$ and $I$ are defined as
\begin{equation}
\varrho_{\Phi}=\left({1\over 2}{\dot{\Phi}^2\over c^2}+V\right){1\over 
c^2},\ \ \ \ \ 
\wp_{\Phi}={1\over 2}{\dot{\Phi}^2\over c^2}-V,\ \ \ \ \ 
I\equiv 1+B{\dot{\Phi}^2\over c^2}.
\label{feqdef}
\end{equation}
The biscalar field equation (\ref{biscleq}) takes the following form:
\begin{equation}
{1\over c^2}\left(1-{{c^2B}\over I^{3/2}}\varrho\right)\ddot{\Phi}+
{3\over c^2}{\dot{a}_0\over a_0}\dot{\Phi}\left(1+{B\over\sqrt{I}}\wp\right)
+V^{\prime}(\Phi)=0.
\label{bsceq}
\end{equation}
In the above equations of motion, we made use of the assumption $n(t,0)=1$ 
to simplify the expressions.  

The derivatives of the metric components w.r.t. $y$ are discontinuous at 
$y=0$ due to the $\delta$-function like brane source there.  From Eqs. 
(\ref{ttein},\ref{iiein}), we obtain the following boundary conditions on 
the first derivatives of $a$ and $n$ at $y=0$:
\begin{equation}
{{[a^{\prime}]_0}\over{a_0b_0}}=-{{8\pi G_5}\over{3c^4}}(\sigma+
\varrho_{\rm tot}c^2),
\label{bc1}
\end{equation}
\begin{equation}
{{[n^{\prime}]_0}\over{n_0b_0}}=-{{8\pi G_5}\over{3c^4}}(\sigma-
3\wp_{\rm tot}-2\varrho_{\rm tot}c^2),
\label{bc2}
\end{equation}
where 
\begin{equation}
\varrho_{\rm tot}\equiv\varrho_{\Phi}+\varrho/\sqrt{I},\ \ \ \ \ \ \ \ \ \ 
\wp_{\rm tot}\equiv\wp_{\Phi}+\sqrt{I}\wp.
\label{bcdef}
\end{equation}
Here, $[F]_0\equiv F(0^+)-F(0^-)$ denotes the jump of a function $F(y)$ 
across $y=0$.  

The effective four-dimensional Friedmann equations on the three-brane 
worldvolume can be obtained \cite{bdl} by taking the jumps and the mean 
values of the above five-dimensional Einstein's equations across $y=0$ 
and then applying the boundary conditions (\ref{bc1},\ref{bc2}) on the 
first derivatives of the metric components.  Here, the mean value of a 
function $F$ across $y=0$ is defined as $\sharp F\sharp\equiv[F(0^+)+
F(0^-)]/2$.  In this paper, we assume that the brane universe is invariant 
under the ${\bf Z}_2$ symmetry, $y\to-y$.  Then, the mean value of the 
first derivative of a function across $y=0$ vanishes.  We also define the 
$y$-coordinate to be proportional to the proper distance along the 
$y$-direction with $b$ being the constant of proportionality, so $b^{\prime}
=0$.  We further assume that the radius of the extra space is stabilized, 
i.e., $\dot{b}=0$, due to some mechanism involving for example a bulk scalar 
field with a stabilizing potential (Cf. Refs. \cite{gw1,gw2}).  Making use of 
these assumptions, we define the $y$-coordinate such that $b=1$.  The 
resulting effective four-dimensional Friedmann equations take the following 
forms:
\begin{equation}
\left({\dot{a}_0\over a_0}\right)^2={{16\pi^2G^2_5}\over{9c^6}}
(\varrho^2_{\rm tot}c^4+2\sigma\varrho_{\rm tot}c^2)+c^2
{a^{\prime\prime}_{R\,0}\over a_0}+{{8\pi G_5}\over{3c^2}}\left(\Lambda+
{{2\pi G_5}\over{3c^4}}\sigma^2\right)-{{kc^2}\over a^2_0},
\label{effrd1}
\end{equation}
\begin{equation}
{\ddot{a}_0\over a_0}=-{{16\pi^2G^2_5}\over{9c^6}}(2\varrho^2_{\rm tot}c^4
+\sigma\varrho_{\rm tot}c^2+3\sigma\wp_{\rm tot}+3\wp_{\rm tot}\varrho_{\rm 
tot}c^2)-c^2{a^{\prime\prime}_{R\,0}\over a_0}+{{16\pi^2G^2_5}\over{9c^6}}
\sigma^2,
\label{effrd2}
\end{equation}
where the subscript $R$ denotes the regular part of a function (note, 
$a^{\prime\prime}$ has a $\delta$-function like singularity at $y=0$).

The $a^{\prime\prime}_{R\,0}$-term (called ``dark radiation'' term) in the 
above Friedmann equations originates from the Weyl tensor of the bulk and thus 
describes the backreaction of the bulk gravitational degrees of freedom on 
the brane \cite{bdl,bdel,ftw,muk,ida}.  This term can be evaluated by solving 
$a^{\prime\prime}_R$ as a function of $y$ from the following equation 
obtained from the Einstein's equations (\ref{ttein},\ref{iiein},\ref{yyein}):
\begin{equation}
3{a^{\prime\prime}_R\over a}+{n^{\prime\prime}_R\over n}=-{{16\pi G_5}\over
{3c^4}}\Lambda,
\label{darkeq}
\end{equation}
along with the following relation obtained from the $(t,y)$-component 
Einstein's equation (\ref{tyein}) with the assumed $\dot{b}=0$ condition:
\begin{equation}
n(t,y)=\lambda(t)\dot{a}(t,y),
\label{anrel}
\end{equation}
where $\lambda(t)$ is an arbitrary function of $t$.  The resulting expression 
is
\begin{equation}
a^{\prime\prime}_R={{\cal C}\over a^3}-{{4\pi G_5}\over{3c^4}}\Lambda a,
\label{darkexp}
\end{equation}
where $\cal C$ is an integration constant.  

To make contact with conventional cosmology having the Hubble parameter 
proportional to $\sqrt{\varrho}$, we assume that $\sigma\gg\varrho_{\rm tot}
c^2, \wp_{\rm tot}$ \cite{cgk,cgs}.  To the leading order, the effective 
Friedmann equations (\ref{effrd1},\ref{effrd2}) along with Eq. (\ref{darkexp}) 
then take the forms
\begin{equation}
\left({\dot{a}_0\over a_0}\right)^2={{32\pi^2G^2_5\sigma}\over{9c^4\sqrt{I}}}
\varrho+{{32\pi^2G^2_5\sigma}\over{9c^6}}\left({1\over 2}{\dot{\Phi}^2\over 
c^2}+V\right)+{{4\pi G_5}\over{3c^2}}\left(\Lambda+{{4\pi G_5}\over{3c^4}}
\sigma^2\right)+{{{\cal C}c^2}\over a^4_0}-{{kc^2}\over a^2_0},
\label{efred1}
\end{equation}
\begin{equation}
{\ddot{a}_0\over a_0}=-{{16\pi^2G^2_5\sigma}\over{9c^4\sqrt{I}}}\left(\varrho
+3I{\wp\over c^2}\right)-{{32\pi^2G^2_5\sigma}\over{9c^6}}\left({\dot{\Phi}^2
\over c^2}-V\right)+{{4\pi G_5}\over{3c^2}}\left(\Lambda+{{4\pi G_5}\over
{3c^4}}\sigma^2\right)-{{{\cal C}c^2}\over a^4_0}.
\label{efred2}
\end{equation}
These effective Friedmann equations for the bimetric brane world cosmology 
have the same forms as the Friedmann equations for the scalar-tensor bimetric 
model of Clayton and Moffat except for the dark radiation term 
${\cal C}c^2/a^4_0$.  

Note, the overdots in the above effective equations denote derivatives 
w.r.t. the time coordinate $t$, with which the matter metric takes the form
\begin{equation}
\hat{g}_{\mu\nu}dx^{\mu}dx^{\nu}=-\left[c^2+B\dot{\Phi}^2\right]dt^2+
a^2_0(t)\gamma_{ij}dx^idx^j,
\label{1matmet}
\end{equation}
and the gravity metric on the brane is given by
\begin{equation}
g_{\mu\nu}dx^{\mu}dx^{\nu}=-c^2dt^2+a^2_0(t)\gamma_{ij}dx^idx^j.
\label{1gravmet}
\end{equation}
Namely, the above effective equations are written in a comoving frame for 
the gravity metric.  As can be seen from these metric expressions, with a 
choice of time coordinate $t$, a graviton travels with a constant speed 
$c_{\rm grav}=c$ and a photon, which is coupled to $\hat{g}_{\mu\nu}$, 
travels with variable speed $c_{\rm ph}=\sqrt{c^2+B\dot{\Phi}^2}=c\sqrt{I}$.  
So, a photon is observed to travel faster than the present day speed in 
this frame, while the biscalar field varies with $t$.  

Since all the matter fields on the brane are coupled to the matter metric 
$\hat{g}_{\mu\nu}$, it would be more natural to consider the comoving frame 
for the matter metric in order to make a connection with standard cosmology.  
By defining the cosmic time $\tau$ of the brane universe in the following way:
\begin{equation}
d\tau^2\equiv (1+B\dot{\Phi}^2/c^2)dt^2,
\label{costime}
\end{equation}
we can bring the matter metric into the following standard comoving frame 
form for the Robertson-Walker metric:
\begin{equation}
\hat{g}_{\mu\nu}dx^{\mu}dx^{\nu}=-c^2d\tau^2+a^2_0(\tau)\gamma_{ij}dx^idx^j.
\label{2matmet}
\end{equation}
In this new frame, the gravity metric (\ref{1gravmet}) takes the form
\begin{equation}
g_{\mu\nu}dx^{\mu}dx^{\nu}=-\left[c^2-B\dot{\Phi}^2\right]d\tau^2+
a^2_0(\tau)\gamma_{ij}dx^idx^j,
\label{2gravmet}
\end{equation}
where the overdot from now on stands for derivative w.r.t. $\tau$.  
So, in the matter metric comoving frame with the time coordinate $\tau$, 
a photon travels with a constant speed $c_{\rm ph}=c$ and a graviton travels 
with a time-variable speed $c_{\rm grav}=\sqrt{c^2-B\dot{\Phi}^2}=
c/\sqrt{I}$.  Note, $I=1/(1-B\dot{\Phi}^2/c^2)$ when the overdot stands for 
derivative w.r.t. $\tau$.  So, a graviton is observed to travel slower 
than the present day speed in this new frame, while $\Phi$ varies 
with $\tau$.  In this new frame, the effective Friedmann equations 
(\ref{efred1},\ref{efred2}) take the following forms:
\begin{eqnarray}
\left({\dot{a}_0\over a_0}\right)^2&=&{{32\pi^2G^2_5\sigma}\over{9c^4I^{3/2}}}
\varrho+{{32\pi^2G^2_5\sigma}\over{9c^6I}}\left({I\over 2}{\dot{\Phi}^2\over 
c^2}+V\right)+{{4\pi G_5}\over{3c^2I}}\left(\Lambda+{{4\pi G_5}\over{3c^4}}
\sigma^2\right)
\cr
& &+{{{\cal C}c^2}\over{a^4_0I}}-{{kc^2}\over{a^2_0I}},
\label{nefffrd1}
\end{eqnarray}
\begin{eqnarray}
{\ddot{a}_0\over a_0}+{1\over 2}{\dot{I}\over I}{\dot{a}_0\over a_0}&=&
-{{16\pi^2G^2_5\sigma}\over{9c^4I^{3/2}}}\left(\varrho+3I{\wp\over 
c^2}\right)-{{32\pi^2G^2_5\sigma}\over{9c^6I}}\left(I{\dot{\Phi}^2
\over c^2}-V\right)
\cr
& &+{{4\pi G_5}\over{3c^2I}}\left(\Lambda+{{4\pi G_5}\over{3c^4}}
\sigma^2\right)-{{{\cal C}c^2}\over{a^4_0I}},
\label{nefffrd2}
\end{eqnarray}
and the biscalar equation (\ref{bsceq}) takes the form
\begin{equation}
{I^2\over c^2}\left(1-{{c^2B}\over I^{3/2}}\varrho\right)\ddot{\Phi}+{{3I}
\over{c^2}}{\dot{a}_0\over a_0}\dot{\Phi}\left(1+{B\over\sqrt{I}}\wp\right)
+V^{\prime}(\Phi)=0.
\label{neffbsc}
\end{equation}

The first Friedmann equation (\ref{nefffrd1}) can be put into the following 
``sum-rule'' form:
\begin{equation}
1+I^{-1}\Omega_k=I^{-3/2}\Omega_{\varrho}+I^{-1}\Omega_{\Phi}+I^{-3/2}
\Omega_{\Lambda_4}+I^{-1}\Omega_{\cal C},
\label{sumrule}
\end{equation}
where the density parameters are defined as
\begin{eqnarray}
\Omega_k&\equiv&{{kc^2}\over{a^2_0H^2}},\ \ \ \ \ \ \ \ 
\Omega_{\varrho}\equiv{{32\pi^2G^2_5\sigma\varrho}\over{9c^4H^2}},
\ \ \ \ \ \ \ \ 
\Omega_{\Phi}\equiv{{32\pi^2G^2_5\sigma\varrho_{\Phi}}\over{9c^4H^2}},
\cr
\Omega_{\Lambda_4}&\equiv&{{32\pi^2G^2_5\sigma\varrho_{\Lambda_4}}\over
{9c^4H^2}},
\ \ \ \ \ \ \ \ \ 
\Omega_{\cal C}={{{\cal C}c^2}\over{a^4_0H^2}}.
\label{defoms}
\end{eqnarray}
Here, $H=\dot{a}_0/a_0$ is the Hubble parameter and $\varrho_{\Lambda_4}=
\Lambda_4c^2_{\rm grav}/(8\pi G_4)={{3c^2\sqrt{I}}\over{8\pi G_5\sigma}}
(\Lambda+{{4\pi G_5}\over{3c^4}}\sigma^2)$ is the vacuum energy density, 
where $\Lambda_4={{4\pi G_5}\over c^4}(\Lambda+{{4\pi G_5}\over{3c^4}}
\sigma^2)$ is the effective four-dimensional cosmological constant.  
Unlike the case of conventional cosmology, the sum rule involves the 
additional factors of $I$.  
From the second Friedmann equation (\ref{nefffrd2}), we obtain the following 
expression for the deceleration parameter:
\begin{equation}
q=-{{a_0\ddot{a}_0}\over \dot{a}^2_0}={\dot{I}\over{2HI}}+{1\over 2}(I^{-3/2}
\Omega_{\varrho}+I^{-1}\Omega_{\Phi})+{{16\pi^2G^2_5\sigma}\over{3c^6H^2}}
(I^{-1/2}\wp+I^{-1}\wp_{\Phi}+2I^{-1/2}\wp_{\Lambda_4})+I^{-1}\Omega_{\cal C},
\label{decelpara}
\end{equation}
where $\wp_{\Lambda_4}=-c^2_{\rm grav}\varrho_{\Lambda_4}=-{{3c^4}\over
{8\pi G_5\sigma\sqrt{I}}}(\Lambda+{{4\pi G_5}\over{3c^4}}\sigma^2)$.  
We consider the special case describing the present day universe having $k=0$, 
$\Lambda_4=0$ and $\wp=0$.  For such case, the sum-rule formula 
(\ref{sumrule}) takes the form:
\begin{equation}
1=I^{-3/2}\Omega_{\varrho}+I^{-1}\Omega_{\Phi}+I^{-1}\Omega_{\cal C}.
\label{prsumrul}
\end{equation}
So, the deceleration parameter (\ref{decelpara}) reduces to
\begin{equation}
q={\dot{I}\over{2HI}}+{1\over 2}+{{16\pi^2G^2_5\sigma}\over{3c^6H^2I}}
\wp_{\Phi}+{1\over{2I}}\Omega_{\cal C}.
\label{prdecel}
\end{equation}
To be consistent with the observational data, the deceleration parameter has 
to be negative.  Unlike the case of conventional cosmology, we have 
additional contribution from the dark radiation term.  A negative value of 
${\cal C}$ helps with achieving negative $q$.  With positive ${\cal C}$, 
more rapid variation of the biscalar field with time is required in order to 
be consistent with the observational data.  

From the effective Friedmann equations (\ref{nefffrd1},\ref{nefffrd2}) in 
the comoving frame for the matter metric, we can read off that the speed 
of a graviton and the effective four-dimensional Newton's constant on the 
brane are respectively given by
\begin{equation}
c_{\rm grav}={c\over\sqrt{I}},\ \ \ \ \ \ \ \ \ \ \ \ \ \ 
G_4={{4\pi G^2_5\sigma}\over{3c^4I^{3/2}}}.
\label{effpara}
\end{equation}
This expression for $c_{\rm grav}$ agrees with the value read off from the 
gravity metric (\ref{2gravmet}).  $G_4$ also varies with time and takes 
smaller value than the present day value while $\dot{\Phi}\neq 0$.  
In terms of these effective four-dimensional parameters, the effective 
Friedmann equations (\ref{nefffrd1},\ref{nefffrd2}) in the comoving frame 
for the matter metric take the forms
\begin{equation}
\left({\dot{a}_0\over a_0}\right)^2={{8\pi G_4}\over 3}\varrho+
{{4\pi G_4}\over{3c^4_{\rm grav}\sqrt{I}}}\dot{\Phi}^2+{c^2_{\rm grav}\over 
3}\Lambda_{\rm eff}+{{{\cal C}c^2_{\rm grav}}\over a^4_0}-{{kc^2_{\rm grav}}
\over a^2_0},
\label{neffrd1}
\end{equation}
\begin{equation}
{\ddot{a}_0\over a_0}+{1\over 2}{\dot{I}\over I}{\dot{a}_0\over a_0}=
-{{4\pi G_4}\over 3}\left(\varrho+3{\wp\over c^2_{\rm grav}}\right)-
{{8\pi G_4}\over{3c^4_{\rm grav}\sqrt{I}}}\dot{\Phi}^2
+{c^2_{\rm grav}\over 3}\Lambda_{\rm eff}-{{{\cal C}c^2_{\rm grav}}\over 
a^4_0},
\label{neffrd2}
\end{equation}
where $\Lambda_{\rm eff}$ is the effective total four-dimensional cosmological 
constant given by
\begin{equation}
\Lambda_{\rm eff}={{4\pi G_5}\over c^4}\left(\Lambda+{{4\pi G_5}\over{3c^4}}
\sigma^2\right)+{{32\pi^2G^2_5\sigma}\over{3c^8}}V(\Phi).
\label{effcos}
\end{equation}
This effective four-dimensional cosmological constant has contribution only 
from $V(\Phi)$, if the brane tension takes the fine-tuned value $\sigma=
\sqrt{-{{3c^4\Lambda}\over{4\pi G_5}}}$ of the RS2 model \cite{rs2}.  

We discuss resolution of various cosmological problems within our bimetric 
model.  First, we consider the horizon problem.  The four-velocity vector 
$V^{\mu}$ of a photon, which is null w.r.t. the matter metric, i.e., 
$\hat{g}_{\mu\nu}V^{\mu}V^{\nu}=0$, is spacelike w.r.t. the gravity metric, 
i.e., $g_{\mu\nu}V^{\mu}V^{\nu}=B(V^{\mu}\partial_{\mu}\Phi)^2>0$ when 
$\partial_{\mu}\Phi\neq 0$ and $B>0$.  So, photons and other matter fields 
propagate outside the lightcone of the gravity metric.  The horizon problem 
is therefore resolved in our bimetric model, provided $\Phi$ varies 
rapid enough with time during an early period of cosmic evolution.  
Furthermore, the problem of unwanted relics such as magnetic monopoles, which 
requires a larger value of the light speed during an early period for its 
resolution in the VSL models, can also be resolved by our bimetric model.  
However, the flatness problem and the cosmological constant problem, which 
require the rapid enough decrease in the speed of a graviton to the present 
day value (in the Friedmann equations) for their resolution in the VSL models, 
cannot be resolved by our bimetric model, since the speed of a graviton takes 
a constant value $c$ in the comoving frame for the gravity metric and takes 
a smaller value than $c$ in the comoving frame for the matter metric, while 
$\dot{\Phi}\neq 0$.  The flatness problem may be resolved by our bimetric 
model, provided the biscalar potential $V(\Phi)$ has a region satisfying the 
slow-roll approximation and thereby the biscalar can act as an inflaton.  
Detailed discussion on resolution of these cosmological problems within the 
VSL brane world cosmologies is given in Refs. \cite{youm,youm1}.  

We comment on the Planck problem of the VSL cosmologies pointed out in Ref. 
\cite{cou}.  When the speed of a graviton and the Newton's constant vary 
with time, so do the Planck mass $m_{pl}=\sqrt{\hbar c_{\rm grav}/G_4}$, 
the Planck length $l_{pl}=\sqrt{\hbar G_4/c^3_{\rm grav}}$ and the Planck 
time $t_{pl}=\sqrt{\hbar G_4/c^5_{grav}}$.  By substituting $c_{\rm grav}$ and 
$G_4$ in Eq. (\ref{effpara}), we see that the Planck mass takes {\it larger} 
value than the present day value, the Planck length remains constant and 
the Planck time takes {\it larger} value, while the biscalar varies with time. 
Since the Planck mass takes larger value, our bimetric model makes the 
hierarchy problem worse.  Furthermore, too much large value of $I$, which 
leads to the value of the Planck time ($\sim c^{-5/2}_{grav}\sim I^{5/4}$) 
larger than $\sim 10^{-20}$ sec would totally mess up the usual standard 
particle physics arguments, e.g., matter dominance over anti-matter.  
Therefore, a judicial choice of the biscalar potential $V(\Phi)$ which 
leads to the value of $I$ not exceeding $\sim 10^{20}$ and therefore the 
speed of light ($\sim I^{1/2}$) not exceeding $\sim 10^{10}$ times the 
present day value is necessary.  This limit on the speed of light during the 
early stage of cosmological evolution may be insufficient for solving the 
cosmological problems.  So, our bimetric model risks the above mentioned 
problem associated with large $t_{pl}$, if it is to solve the cosmological 
problems.  However, since the Planck density ($\sim m_{pl}/l^3_{pl}\sim 
I^{1/2}$) increases for our bimetric model, the Planck density problem may 
be resolved.

\end{document}